\documentclass{article}

\usepackage{arxiv}

\usepackage[utf8]{inputenc} 
\usepackage[T1]{fontenc}    
\usepackage{booktabs}       
\usepackage{amsfonts}       
\usepackage{nicefrac}       
\usepackage{microtype}      
\usepackage{lipsum}
\usepackage{cite}
\usepackage{amsmath,amssymb,amsfonts}
\usepackage{algorithmic}
\usepackage{graphicx}
\usepackage{textcomp}
\usepackage{multirow}
\usepackage{multicol}
\usepackage{xcolor}
\newcommand{\shield}{Entropy-Shield}
\def\BibTeX{{\rm B\kern-.05em{\sc i\kern-.025em b}\kern-.08em
    T\kern-.1667em\lower.7ex\hbox{E}\kern-.125emX}}
\begin{document}

\title{Adversarial Learning Inspired Emerging Side-Channel Attacks and Defenses\\
}

\author{Abhijitt Dhavlle \\
Electrical and Computer Engineering \\
\textit{George Mason University}\\
Fairfax, USA. \\
adhavlle@gmu.edu}

\maketitle

\begin{abstract}
Evolving attacks on the vulnerabilities of the computing systems demand novel defense strategies to keep pace with newer attacks. This report discusses previous works on side-channel attacks (SCAs) and defenses for cache-targeted and physical proximity attacks. We then discuss the proposed \shield\ as a defense against timing SCAs, and explain how we can extend the same to hardware-based implementations of crypto applications as ``Entropy-Shield for FPGA". We then discuss why we want to build newer attacks with the hope of coming up with better defense strategies. 
\end{abstract}


\section{Introduction} \label{sec:intro}
The hardware security domain in recent years has experienced a
plethora of threats Side-Channel Attacks \cite{Yarom_usenix_'14,Gruss_dimva_'16}, Malware attacks \cite{Dhavlle_DATE'21, Dhavlle_ISCAS'21, Meraj_ISCAS'21, SMPD_DAC'19, Sanket_CASES'19, Sanket_ICMLA'19, Sanket_ICTAI'19}, Hardware Trojan attacks \cite{Meraj_AsianHOST'20}, reverse engineering threats \cite{Kolhe_GLSVLSI'19, Kolhe_ICCAD'19, Hassan_ISQED'20} and so on. Among multiple threats, the side-channel attacks (SCAs) is one of the pivotal threats due to it's capability to exploit the design despite being introduced in the market post-validation. These SCAs function by exploiting
the side-channels, which invariably leak important data during
an application’s execution. The information leaked through side-channels are inherent characteristics of the system and are often unintentional. 
The side-channel information can be device's
microarchitectural or physical information such as 
power consumption, thermal maps, the timing of the operations, acoustics, 
and cache access traces. Some of the attacks are reported in \cite{report_1, report_3, report_4} with vulnerabilities in Linux been reported in \cite{report_5}. The work in \cite{report_2} lists and reports the common vulnerabilities in the computing systems discovered so far, while some newer threats to a CPU have been reported and described in \cite{report_6}. It is quite evident by looking at the mentioned reports that the side-channel attacks have proven a major threat to system security, and developing novel mechanisms to mitigate them becomes inevitable. 

Snooping on co-located programs' cache trace is performed by cache-based SCAs and is prominently explored in the domain of SCAs. A part of my thesis work focuses on timing-based cache attacks, which exploit the time required to flush/reload an instruction from the cache subsystem as a covert channel\cite{Abhijitt_isqed'20,Dhavlle_TCAD'21}. Some of the popular cache targeted SCAs are the Flush+Reload\cite{Yarom_usenix_'14} and Flush+Flush attack\cite{Gruss_dimva_'16}.
Intercepting secret information based on the study of power signature is a subdivision of SCAs where power consumption information serves as a covert channel leaking crucial information about the executed operations. 
Such physical and cache-based SCAs are known to be a significant threat to cryptosystems such as AES (Advanced Encryption Standard) and RSA and can reveal the encryption key efficiently. Yet these attacks are not limited to the cryptographic algorithms, and they can be extended to snoop on other applications as well. 
One can argue to have a 
solution that can simply shut the covert channels and hence prevent such threats in the first place. It is not as straightforward as it seems to be because shutting down covert channels that originated due to vulnerability is not feasible due to the cost and performance constraints. Hence, in this report, we introduce the reader to primarily cache-targeted and physical SCAs, discuss the proposed solution, existing works, and how it compares with our solution, along with current ongoing and future projects. 

The rest of this report is organized as follows: Section \ref{sec:cache_based} enumerates and describes the existing works on cache-targeted attacks and defenses, and our proposed defense; Section \ref{sec:physical_based} mentions previous works on physical SCAs and the ongoing research in the domain; Section \ref{sec:conclusion} concludes this report with the anticipated results and contributions considering the current ongoing research.



\section{Cache-Targeted Side-Channel Attacks/Defenses} \label{sec:cache_based}

Cache-targeted attacks primarily rely on the timing information to successfully probe the intended process/data. The idea is to co-locate 
with a victim and snoop on its private data over the covert channel. 
Intel CPUs use \textit{cflush} instruction to evict a specified cache line from the cache subsystem. Intel uses inclusive caches, meaning a line evicted from the cache using the \textit{cflush} command does evict it from all the levels of cache. The block diagram of the cache subsystem is shown in Figure \ref{fig:cache_system}(b). The attackers time the \textit{cflush} instruction after evicting a cache line belonging to the victim. Depending on the access time, 
the attacker deduces if the victim accessed the data. Flush+Flush\cite{Gruss_dimva_'16} and Flush+Reload\cite{Yarom_usenix_'14} are some of the most popular cache-targeted SCAs exploiting the timing information to steal private data. 

In this section, we discuss the existing cache-targeted SCAs, defenses, and our proposed \shield\ as a defense against cache-targeted attacks. 
\subsection{Existing Attacks Targeting Cache} 
\textbf{Flush+Flush}: Flush+Flush\cite{Gruss_dimva_'16} is a passive type of an attack unlike its sibling, Flush+Reload\cite{Yarom_usenix_'14}, in the sense that it does not reload the data again to check a hit/miss. This attack depends on the time it takes for the \textit{cflush} instruction to complete. Because the system has to flush all the copies of the flushed data, it requires some time to do so, and hence, flushing a data that is present in the cache takes more time as against missing data. Hence, Flush+Flush can silently attack a victim and steal its secret data. 

\textbf{Prime+Probe}: The Prime+Probe\cite{Liu_prime_probe} attack targets the LLC. The attacker 'primes' or prepares its cache by loading a chunk of data. It then waits for the victim to execute. If the victim's data is mapped to the same area as the attacker's data, the attacker data is evicted given the replacement policy. When the attacker reaccesses its data, it deduces if the victim's data was present, depending on whether the 'probe' was a hit or a miss.

\begin{figure}[tb!]
    \centering
    \includegraphics[width=1\textwidth]{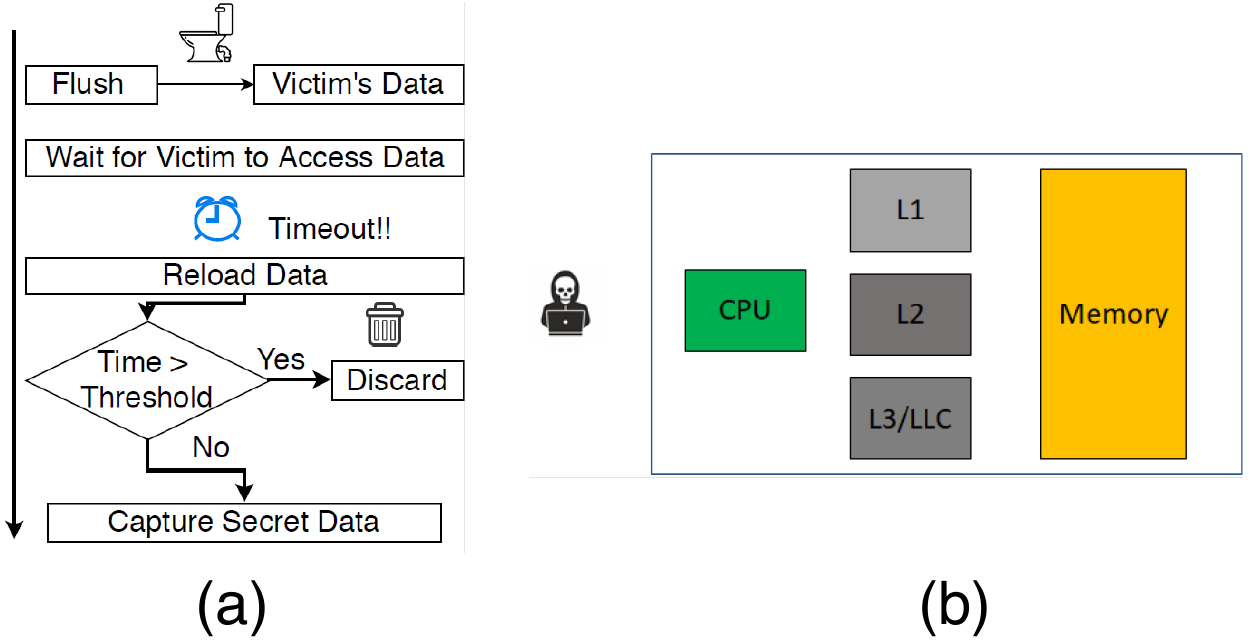}
    \caption{(a)Flush+Reload Attack Principle; (b)Block Diagram of the Cache Sub-system}
     \label{fig:cache_system}
\end{figure}

\textbf{}
\textbf{Flush+Reload}: 
A quick background on the Flush+Reload attack is discussed here. Figure \ref{fig:cache_system}(a) demonstrates the attack process: 
\begin{enumerate}
    \item The attacker flushes the victim's cache line, which corresponds to a line/function in victim's code.
    \item The attacker waits for the victim to access its data.
    \item The attacker again reloads the same data and times this access.
    \item If the time happens to be greater than the threshold, it means the victim did not access the data. This event corresponds to a cache miss.
    \item If the access time is less than the threshold, the victim accessed the data. This corresponds to a cache hit.
\end{enumerate}


\subsection{Existing Defenses for Cache-Targeted SCAs} 
The existing defenses against cache-targeted SCAs can be broadly categorized into five categories, as discussed below. 

\subsubsection{Randomization based} 
Cases where modifying the hardware is not feasible, randomizing the memory accesses has been proposed in \cite{Wang_isca_'07}. The work in \cite{He_micro_'17} proposed the use of random memory-to-cache mappings to make the attack difficult. A permutation table for each process causes dynamic memory address to cache set maps. The attacker relies on evicting 
a specific cache line to perform the attack, dynamic memory address 
mapping prevents the attacker from evicting the targeted cache line. 
This comes for a price though, as maintaining the table during runtime 
affects performance. 
A dynamic random control flow by exploiting software diversity is introduced in 
\cite{Crane_ndsss"15}. 
The main idea is to transform the program trace of each application into a unique trace. The technique dynamically randomizes the control flow of the program during runtime. Replicas of functions within the code are created, and the dynamic program flow chooses one of them in runtime to randomize the overall control flow. 

\subsubsection{Partition based}
DAWG by Kiriansky\cite{Kiriansky_micro'18} proposes a Dynamically Allocated Way Guard (DAWG) to protect way partitions in a set-associative cache. Secure isolation is provided by giving a notion of protected domains within the cache. DAWG is capable of isolating cache miss, hit, and other metadata across the protected domains. DAWG can protect against attacks that depend on timing behaviors. The defense mechanism can be extended to branch history tables or the TLBs. However, DAWG would need supplemental techniques to block covert channels, and this defense leads to performance overhead. Another partitioning based technique is proposed in \cite{He_micro_'17}, which promises to protect sensitive data from attack by reserving dedicated cache sets. The sensitive process, in this way, would map to the protected cache area, thereby reducing the interference caused by other programs. But, this method requires the intervention of the operating system to separate cache into a reserved and non-reserved area, all of which must be non-overlapping. This brings overhead and inefficient cache utilization, along with security. 

\subsubsection{Intel Cache Allocation Technology (CAT)}
Work in \cite{Liu_hpca'16} utilized Intel's CAT technology to protect SCAs on shared last-level cache. CAT is a way partitioning mechanism that ensures efficient utilization of the cache space/occupancy. Another work in \cite{Dong_usenix'18} proposes to use CAT in addition to the mechanism that can partition the LLC to thwart SCAs. CAT cannot by itself securely partition the cache. The attack launches on a compromised operating system. Hence, by preventing cache line sharing with applications or OS, the authors propose to defend against SCAs. 

\subsubsection{Detection based defenses}
Cloudradar\cite{Zhang_raid_'16} proposes to detect SCAs using hardware performance counters (HPCs). Specialized cores are employed to detect SCAs based on either signature detection technique or anomaly-based detection established on HPCs. The system constantly monitors the HPCs. These traces are then compared with pre-stored attack signatures to detect the presence of an attack. Another technique used is the anomaly detection technique, which alerts the system based on the drift of the program flow from its normal flow. Stealthmem\cite{Kim_usenix'12} provides a system-level protection against cache SCAs. It protects the cache from unauthorized accesses by providing a set of locked cache lines per core. These cache lines are never evicted. Authors in \cite{Shi_dsnw'11} develop a technique, titled as dynamic cache coloring. This technique cautions the virtual machine when a security-sensitive operation is being processed. This swaps the data to a safe, isolated cache line that has limited access, thus preventing the attack.  

\subsubsection{SGX Enclave based}
Varys\cite{Oleksenko_usenix'18} proposes to protect the system from timing based and page-table attacks. The main idea is to prevent shared resources. During the execution of secure-sensitive operations in the enclave, strict reservation of physical cores is enforced. 
These previous works have been summarized in Table \ref{tbl:cache_summary}.

\begin{table*}[!htb]
\centering
\caption{Cache Targeted Defenses Summarized} 
\label{tbl:cache_summary} 
\scalebox{0.63}{
\begin{tabular}{|c|c|}
\hline
Title/Reference & Technique \\
\hline
``New Cache" \cite{Wang_isca_'07} & Randomize memory maps. \\
\hline
``How secure is your cache?" \cite{He_micro_'17} & Use permutation tables to randomize cache mapping to make it harder to evict specific targeted cache line. \\
\hline
``How secure is your cache?" \cite{He_micro_'17} & Sensitive process access dedicated non-overlapping cache set. \\
\hline
``Varys" \cite{Oleksenko_usenix'18} & Protects programs in SGX enclaves from cache attacks.Strict reservation of physical cores to threads.\\
\hline 
``Cloudradar" \cite{Zhang_raid_'16} & Cores  (processors)  equipped  with specialized signature detection. Detects  SCAs  based  on  the  hardware  performance  counters(HPCs). \\

\hline
``Stealthmem" \cite{Kim_usenix'12} & Protects cache  from unauthorized access by creating set  of  locked  cache lines  per  core. Lines are never evicted  from  the  cache.  
\\

\hline

``Dynamic Cache Coloring" \cite{Shi_dsnw'11} & Notifies VM when application executing secure-sensitive operations.\ Swaps associated data to safe and isolated cache line, limiting its access. \\

\hline
``Partition Locked (PL) Cache" \cite{Wang_isca_'07} & Protected cache line cannot be evicted by an attacker’s miss. \\

\hline
``Non-monopolizable Cache" \cite{Domnister_nomo'12} & Sensitive process assigned reserved ways in each cache set. Disallow attacker from occupying whole cache set. \\
\hline
\end{tabular}
}
\end{table*}

\subsection{Proposed Entropy-Shield}
In this subsection, we discuss the vulnerability in the GnuPG RSA encryption algorithm and propose a defense mechanism to thwart attacks. 
\subsubsection{Vulnerability in RSA}
The vulnerabilities in the existing RSA algorithms is described in \cite{Yarom_usenix_'14} and also demonstrates how Flush+Reload can successfully retrieve all the secret key bits. Referring to Table \ref{tbl:rsa}, bit `0' in a secret key would translate to RSA executing a square operation followed by a modulo operation. Likewise, square-modulo-multiply-modulo operations are called for a bit `1'. Table \ref{tbl:rsa_sequence} shows the sequence of operations corresponding to different key bit combinations. By observing the functions calls to the square, reduce or modulo lines of code, the attacker can deduce the secret key bits. Not all the bits need to be retrieved at once, but the attacker has to execute the attack for several hundreds of thousands iteratively to guess the correct key. 

\subsubsection{Entropy-Shield} 
\begin{figure*}[tb!]
    \centering
    \includegraphics[width=1\textwidth]{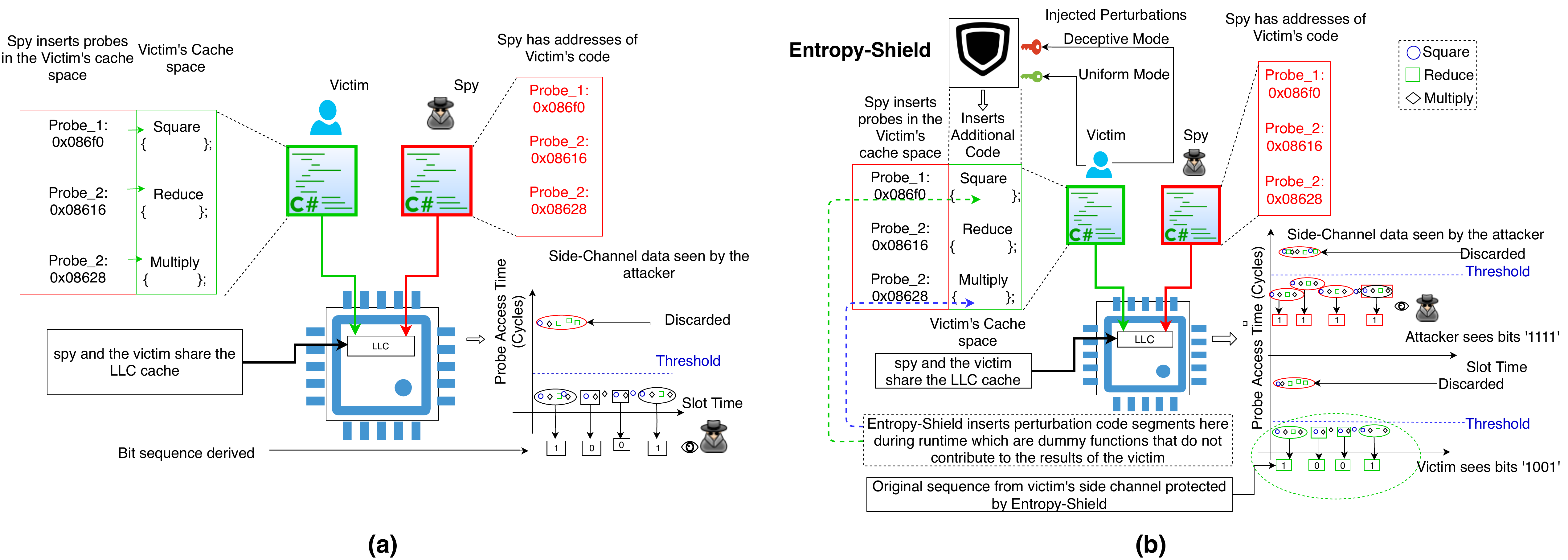}
    \caption{(a) Traditional side-channel attack on encryption algorithm where the data leaked via covert channel is accessible to the attacker; (b) Victim wrapped with Entropy-Shield that injects perturbation during run-time to perturb the sensitive information leaked thereby making SCAs laborious and time-consuming. *Only Uniform mode results have been shown* \cite{Abhijitt_isqed'20}}
     \label{fig:full_diagram}
\end{figure*}

Earlier, we discussed different attacks and defense strategies. With the defenses discussed, it is evident that some of them require software level changes to the operating system or the hypervisor, while others require hardware modifications, causing inefficient utilization of the cache or severe overheads. Hence, we propose \shield\ as a defense against timing-based SCAs, shown in Figure \ref{fig:full_diagram}. 
In Figure \ref{fig:full_diagram}(a), we can see that the attacker can sniff the covert channel and hence, retrieve secure-sensitive data. With \shield\, the attacker is misled to an incorrect key. This is achieved using fake/dummy calls to the square, reduce, or multiply functions. We call this as adding perturbations to the signal/channel observed by the attacker. For example, if a bit `0' is processed by the victim, the \shield\ calls dummy Multiply-Modulo operations following the original Square-Modulo operations, thus giving the notion of bit `1' processed by the victim when bit `0' was processed. Refer to Table \ref{tbl:fake_sequence} for the fake operations corresponding to key bits. 
Referring to Figure \ref{fig:full_diagram}(b), with \shield\, dummy operations are called to mislead the attacker. The \shield\ offers \textit{Uniform} and \textit{Deceptive} mode of operation. \textit{Uniform} mode adds maximum noise to the channel by flipping all 0's to 1's, while \textit{Deceptive} mode cognitively perturbs the sequence by randomly calling dummy operations, the position of the bits perturbed change every run. The results of the \shield\ from \cite{Abhijitt_isqed'20} is presented in in Tables \ref{tbl:shield_results_uniform}
and \ref{tbl:shield_results_deceptive}. 

\begin{table}[!htb]
\centering
\caption{Operations Corresponding to Key Bits} 
\label{tbl:rsa} 
\scalebox{1}{
\begin{tabular}{|c|c|}
\hline
Secret Key Bit & Corresponding Operations \\
\hline
Bit `0' & Square-Modulo \\        
\hline
Bit `1' & Square-Modulo-Multiply-Modulo \\ 
\hline
\end{tabular}
}
\end{table}
\begin{table}[!htb]
\centering
\caption{Sequence of Key Bits to Corresponding Operations} 
\label{tbl:rsa_sequence} 
\scalebox{1.2}{
\begin{tabular}{|c|c|}
\hline
Bit Sequence & Corresponding Operations \\
\hline
``0010" &  SR-SR-SRMR-SR\\        
\hline
``1100" &  SRMR-SRMR-SR-SR\\
\hline
\end{tabular}
}
\end{table}

\begin{table}[!htb]
\centering
\caption{Sequence of Key Bits to Corresponding Operations} 
\label{tbl:fake_sequence} 
\scalebox{1}{
\begin{tabular}{|c|c|}
\hline
Secret Key Bit & Corresponding Fake Operations \\
\hline
Bit `0' &  Square-Modulo\\        
\hline
Bit `1'  &  Square-Modulo-\textcolor{red}{Multiply-Modulo} \\
\hline
\end{tabular}
 } 
\end{table}

\begin{table*}[htb!]
\centering
\caption{Key as visible to the attacker and the victim with \shield\ - Uniform mode of operation} 
\label{tbl:shield_results_uniform} 
\scalebox{1}{
\begin{tabular}{|c||c|c|c|c|c|}
\hline
Attack Type & Encryption & Key  &  Original Key & Victim seen key & Key seen by the attacker \\

\hline

\multirow{2}{6em}{Flush+Reload}  & RSA-RSA & key\_1 & 0FCFFF & 0FCFFF & FFFFFF \\        
\cline{2-6}
& DSA-Elgamal & key\_2 & 587BFA & 587BFA &  FFFFFF\\
\cline{1-6}
\multirow{2}{6em}{Flush+Flush }& RSA-RSA & key\_3 & 54FF0B & 54FF0B & FFFFFF \\
\cline{2-6}
& DSA-Elgamal & key\_4 & 89DE00 & 89DE00 & FFFFFF \\
\hline

\end{tabular} 
}
\end{table*}

\begin{table*}[htb!]
\centering
\caption{Key as visible to the attacker and the victim with \shield\ - Deceptive mode of operation} 
\label{tbl:shield_results_deceptive} 
\scalebox{1}{
\begin{tabular}{|c||c|c|c|c||c|c|}
\hline
Attack Type & Encryption & Key  &  Original Key & Victim seen key & \multicolumn{2}{c|}{Key seen by the attacker}  \\
\hline
 & & & & & Iteration 1 & Iteration 100th \\
 \hline
\multirow{2}{6em}{Flush+Reload}  & RSA-RSA & key\_1 & 0FCFFF & 0FCFFF & \textcolor{red}{5}F\textcolor{red}{D}FFF & \textcolor{red}{0}F\textcolor{red}{E}FFF \\        
\cline{2-7}
& DSA-Elgamal & key\_2 & 587BFA & 587BFA &  5\textcolor{red}{9F}BF\textcolor{red}{B} & \textcolor{red}{7}8\textcolor{red}{FF}F\textcolor{red}{E} \\
\cline{1-7}
\multirow{2}{6em}{Flush+Flush }& RSA-RSA & key\_3 & 54FF0B & 54FF0B & \textcolor{red}{75}FF\textcolor{red}{1}B & 5\textcolor{red}{5}FF\textcolor{red}{1F} \\
\cline{2-7}
& DSA-Elgamal & key\_4 & 89DE00 & 89DE00 & \textcolor{red}{CB}D\textcolor{red}{F}0\textcolor{red}{2} & 8\textcolor{red}{B}D\textcolor{red}{F}0\textcolor{red}{1} \\
\hline

\end{tabular}
} 
\end{table*}

\section{Physical Side-Channel Attacks} \label{sec:physical_based}
Physical side-channel attacks are those that require proximity to a victim device - an FPGA or a CPU. The attack principle is to target data-dependent power traces/signatures of the victim device. Crypto algorithm like the AES demonstrates data-dependent power consumption which serves as a covert channel. The attacker attaches a probe to measure the power traces by executing the application iteratively. By applying methods such as Correlation Power Analysis (CPA), Differential Power Analysis (DPA), or the Simple Power Analysis (SPA), the attacker can reveal the entire secret key. This data-dependent behavior can be extended to other non-cryptographic applications as well to snoop on the private data. 
In this section, we discuss the physical side-channel attacks, those that require proximity to the target hardware, as well as the remote physical side-channel attacks. 
We review the existing attacks, defenses, and our ongoing work in this domain. 

\subsection{Existing Power-based Physical Side-Channel Attacks} 
Work in \cite{Zhao_remote_fpga'18} exploits a new vulnerability in the power-sharing within FPGA to snoop on other entity's power consumption. The power rails are shared across all the components. The attack employs ring oscillator (RO) -based design to convert voltage drops to figures, representing power consumption of co-located applications on the same FPGA. The attack describes the successful recovery of the RSA key. Authors in \cite{Ramesh_without_access'18} bring to light yet another type of vulnerability. The work discusses how an RO-based spy can read the relative delay in the wire, a delay caused by logic values on the adjacent victim wire. The crosstalk between two wires - transmitter (victim) and the receiver (spy) - is exploited in this type of attack. Hardware designs have a strict power budget, which causes one component to power throttle while another power-hungry component is operating. This power budget has been exploited as a covert channel in \cite{Khatamifard_powert_19}. The source entity is assumed to have access to sensitive data while the sink has access to a third party application that reads the data. Using power viruses, the attack causes other known (spy) entity to power throttle, thus utilizing this reduced performance factor to correspond to logical values `0' or `1'. Simple Power Analyis (SPA)\cite{SPA'07,Mangard_spa'03} , Difference-of-Means Power Analysis (DPA)\cite{Lo_dpa_cpa'17} and Correlation Power Analysis(CPA)\cite{Lo_dpa_cpa'17} are analysis methodologies used to extract secret data from the observed power traces of an application. Simple Power Analysis is the most simple type of them. It requires manual observation of the power traces to extract the key. DPA requires the collection of a large number of traces. These traces are processed statistically to find the mean of differences observed in the traces, to derive the secret information. While in CPA, the correlation between the measured power traces and the guesses (from a hypothetical power model) is calculated to conclude the secret key.  

\subsection{Existing Defenses} 
Nele Mentens in \cite{Mentens_randomization'17} proposed to utilize hardware components of the FPGA to thwart attacks. The reconfigurations to prevent leakage are performed in runtime. The author discusses the addition of random noise, irregular clock cycles, and scrambling S-Box in the hardware to thwart attacks. Work in \cite{Yu_codes'07} proposed an intelligent place-and-route technique to perform symmetrical routing. This approach is explained to defense against power analysis SCAs. Authors in \cite{Kocher_crypto'99} explain the algorithmic countermeasures that minimize the correlation between power and processed data. 

\subsection{Ongoing work in Physical Side-Channel Attacks domain}
The ongoing work in physical SCAs is to research on methods to secure FPGA implementation of crypto algorithms from attacks. We are working on CPA type attacks on power traces collected during the execution of the AES application. Once the attack is successful, securing the FPGA with approximate modules is what we plan to focus on in the future. Post power trace collection on the secure design, we would attack with CPA again to evaluate the resilience of the new security measure implemented.

\section{Research Progress:}
As described in the previous sections, I have been working and publishing papers on Malware Detection \cite{SMPD_DAC'19}, Side-Channels Analysis \cite{Abhijitt_isqed'20, Dhavlle_TCAD'21}, Hardware-based Trojan Attack and Detection \cite{Meraj_AsianHOST'20}, and Survey-based papers \cite{Brasser_CASES'18, Dhavlle_IGSC'20} published to conferences and journals. I intend to dive deeper into more of SCA based attacks and defenses in future and contribute my work to top tier conferences and journals. 

\section{Conclusion} \label{sec:conclusion}
Utilizing traditional randomization is not the best solution -given the overhead, performance, and design requiring significant hardware modifications. As a panacea, we discussed our proposed \shield\ that would offer security with significantly less overhead. Our \shield\ incurs less overhead (2$\times$). With the positive results achieved so far, we would like to extend this work further in the hardware domain, thus enabling cognitive randomization based defense for FPGAs, which would secure them from side-channel attacks. Furthermore, working on approximate modules for AES, adversarial learning, and one-shot learning with neural networks, we want to contribute to the community by building attacks that would help us evaluate our defenses more accurately.

\bibliographystyle{IEEEtran}
\bibliography{rqe_main}
\end{document}